\documentclass[conference]{IEEEtran}
\IEEEoverridecommandlockouts

\usepackage{amsmath,amssymb,amsfonts}
\usepackage{graphicx}
\usepackage{booktabs}
\usepackage{bm}
\usepackage{siunitx}
\usepackage{comment}
\usepackage[ruled,vlined]{algorithm2e}
\usepackage[backend=biber,style=ieee,
minbibnames=6,
maxbibnames=6,
doi=false,isbn=false,url=false,eprint=false
]{biblatex}
\defbibheading{bibliography}[\refname]{}

\DeclareSourcemap{
	\maps[datatype=bibtex, overwrite=true]{
		\map{
		    \step[fieldsource=booktitle,
			match=\regexp{.*EUSIPCO.*},
			replace={Proc. EUSIPCO}]
			\step[fieldsource=booktitle,
			match=\regexp{.*CVPR.*},
			replace={Proc. CVPR}]
			\step[fieldsource=booktitle,
			match=\regexp{.*Interspeech.*},
			replace={Proc. Interspeech}]
			\step[fieldsource=booktitle,
			match=\regexp{.*ICASSP.*},
			replace={Proc. ICASSP}]
			\step[fieldsource=booktitle,
			match=\regexp{.*ICLR.*},
			replace={Proc. ICLR}]
			\step[fieldsource=booktitle,
			match=\regexp{.*ICCV.*},
			replace={Proc. ICCV}]
			\step[fieldsource=booktitle,
			match=\regexp{.*ICML.*},
			replace={Proc. ICML}]
			\step[fieldsource=booktitle,
			match=\regexp{.*ASRU.*},
			replace={Proc. ASRU}]
			\step[fieldsource=booktitle,
			match=\regexp{.*SLT.*},
			replace={Proc. SLT}]
			\step[fieldsource=booktitle,
			match=\regexp{.*SSW.*},
			replace={Proc. SSW}]
			\step[fieldsource=booktitle,
			match=\regexp{.*WASPAA.*},
			replace={Proc. WASPAA}]
			\step[fieldsource=booktitle,
			match=\regexp{.*IJCNN.*},
			replace={Proc. IJCNN}]
            \step[fieldsource=booktitle,
            match=\regexp{.*Detection.*and.*Classification.*of.*Acoustic.*Scenes.*and.*Events.*Workshop.*},
			replace={Proc. DCASE}]
            \step[fieldsource=booktitle,
			match=\regexp{.*MLSP.*},
			replace={Proc. MLSP}]
            \step[fieldsource=booktitle,
			match=\regexp{.*ECCV.*},
			replace={Proc. ECCV}]
            \step[fieldsource=journal,
			match=\regexp{.*NeurIPS.*},
			replace={Advances in NuerIPS}]
            \step[fieldsource=journal,
			match=\regexp{.*TASLP.*},
			replace={IEEE/ACM TASLP}]
            \step[fieldsource=journal,
			match=\regexp{.*J-STSP.*},
			replace={IEEE J-STSP}]                
			\step[fieldsource=series,
			match=\regexp{.+},
			replace={{}}]
			\step[fieldsource=editor,
			match=\regexp{.+},
			replace={{}}]
			\step[fieldsource=publisher,
			match=\regexp{.+},
			replace={{}}]
			\step[fieldsource=month,
			match=\regexp{.+},
			replace={{}}]
			\step[fieldsource=location,
			match=\regexp{.+},
			replace={{}}]
			\step[fieldsource=address,
			match=\regexp{.+},
			replace={{}}]
			\step[fieldsource=organization,
			match=\regexp{.+},
			replace={{}}]
		}
	}
}

\begin{document}

\title{
    Improvements of Discriminative Feature Space Training for Anomalous Sound Detection in Unlabeled Conditions

\thanks{This paper was partly supported by a project, JPNP20006, commissioned by NEDO, and JSPS KAKENHI Grant Number JP20H00102.}
}

\author{\IEEEauthorblockN{Takuya Fujimura} 
\IEEEauthorblockA{\textit{Graduate School of Informatics} \\
\textit{Nagoya University}\\
Nagoya, Japan \\
fujimura.takuya@g.sp.m.is.nagoya-u.ac.jp}
\and
\IEEEauthorblockN{Ibuki Kuroyanagi} 
\IEEEauthorblockA{\textit{Graduate School of Informatics} \\
\textit{Nagoya University}\\
Nagoya, Japan \\
kuroyanagi.ibuki@g.sp.m.is.nagoya-u.ac.jp}
\and
\IEEEauthorblockN{Tomoki Toda} 
\IEEEauthorblockA{\textit{Information Technology Center} \\
\textit{Nagoya University}\\
Nagoya, Japan \\
tomoki@icts.nagoya-u.ac.jp}
}

\maketitle

\begin{abstract}
In anomalous sound detection, the discriminative method has demonstrated superior performance.
This approach constructs a discriminative feature space through the classification of the meta-information labels for normal sounds.
This feature space reflects the differences in machine sounds and effectively captures anomalous sounds.
However, its performance significantly degrades when the meta-information labels are missing.
In this paper, we improve the performance of a discriminative method under unlabeled conditions by two approaches.
First, we enhance the feature extractor to perform better under unlabeled conditions.
Our enhanced feature extractor utilizes multi-resolution spectrograms with a new training strategy.
Second, we propose various pseudo-labeling methods to effectively train the feature extractor.
The experimental evaluations show that the proposed feature extractor and pseudo-labeling methods significantly improve performance under unlabeled conditions.
\end{abstract}


\vspace{-5pt}
\section{Introduction}
\vspace{-2pt}
\label{sec:intro}
Anomalous sound detection~(ASD) is used in machine condition monitoring to automatically detect abnormal machine behavior based on their operational machine sounds~\cite{Koizumi_DCASE2020_01,kawaguchi2021description,dohi2022description,Dohi_arXiv2023_01,Nishida_arXiv2024_01}. 
Since it is not feasible to collect data for all possible types of anomalous behavior, ASD systems are often developed using only normal sound data.
Therefore, we detect anomalies based on the degree of deviation from the normal sound distribution.

Most state-of-the-art ASD methods are based on a discriminative approach~\cite{Wilkinghoff2023design,wilkinghoff2024self,kuroyanagi2022improvement,liu2022anomalous,han2024exploring}.
This approach consists of two steps.
First, a feature extractor is trained to classify various types of normal sounds utilizing annotated labels, such as machine types and operational parameters, developing a discriminative feature space.
Then, it calculates anomaly scores based on the deviations from the normal sound distribution within the discriminative feature space.
This discriminative approach has shown superior performance compared to a generative approach, which directly models the distribution of normal sounds in the audio feature space~\cite{Koizumi_DCASE2020_01,suefusa2020anomalous,dohi2021flow,dohi2022disentangling,guan2023time}.
This is because the discriminative training can construct a feature space that emphasizes differences in machine sounds while ignoring noise~\cite{wilkinghoff2023angular}.
However, the performance of the discriminative approach heavily depends on the availability of labels.
In~\cite{wilkinghoff2023angular}, it has been shown that the performance significantly degrades when some information in the labels is unavailable.
This poses a practical challenge, as collecting labels is often costly or not always available.

The goal of this work is to improve the performance of the discriminative approach under unlabeled conditions.
We approach this goal in two ways.
First, to achieve better performance under the unlabeled conditions, we enhance the discriminative feature extractor.
Specifically, we propose a feature extractor that utilizes multi-resolution spectrograms and its training strategy.
Second, we effectively train the feature extractor using pseudo-labels.
We generate them in various feature spaces.
The experimental results demonstrate that 
1) the enhanced feature extractors improve the performance with and without the labels, and
2) the pseudo-labeling methods significantly improves the performance in the unlabeled conditions.
Also, 3) we analyze the differences in effectiveness among the proposed pseudo-labeling methods.
The codes are available at our repository\footnote{\texttt{https://github.com/TakuyaFujimura/unlabeledasd}}.

\vspace{-1pt}
\section{State-of-the-art method \\ under the labeled conditions}
\vspace{-1pt}
\label{sec:related}
In this section, we describe the state-of-the-art discriminative method under labeled conditions~\cite{wilkinghoff2021sub,Wilkinghoff2023design,wilkinghoff2024self}.
These works are based on the DCASE Challenge Task2~\cite{Koizumi_DCASE2020_01,kawaguchi2021description,dohi2022description,Dohi_arXiv2023_01,Nishida_arXiv2024_01}, and the labels of this task include machine types, domains (source or target domains), and attributes (operational paramters).
Most of the training data are from the source domain, and only a few samples are from the target domain.
The two domains do not contain samples with the same attribute (i.e., the recording conditions differ between domains).

\vspace{-4pt}
\subsection{Structure of the feature extractor}
\vspace{-2pt}
The feature extractor receives two types of input features: an amplitude spectrum and an amplitude spectrogram~\cite{Wilkinghoff2023design}.
These input features are independently processed by two neural networks.
Each network extracts a $D$-dimensional feature from each input feature and obtains $\bm{z}^{\rm cat}_i$ for the $i$-th audio signal $\bm{x}_i\in\mathbb{R}^T$ as follows:
\begin{align}
    \vspace{-6pt}
    \bm{z}^{\rm cat}_i&=[\bm{z}^{(1)}_i, \bm{z}^{(2)}_i]\in\mathbb{R}^{2D},
    \vspace{-6pt}
\end{align}
where $T$ is the length of the time-domain audio signal and $\bm{z}^{(m)}_i=g_m(f_m(\bm{x}_i))$.
$f_1(\cdot)$ and $f_2(\cdot)$ apply discrete Fourier transform~(DFT) and short-time Fourier transform~(STFT) to the audio signal, respectively.
$g_1(\cdot)$ and $g_2(\cdot)$ are neural networks.
The combination of the spectrum and the spectrogram compensates for each other's information and has been shown to improve the performance~\cite{Wilkinghoff2023design}.

\subsection{Training of the feature extractor}
The feature extractor is trained with an angular margin loss such as ArcFace~\cite{deng2019arcface}, AdaCos~\cite{zhang2019adacos}, and Sub-cluster AdaCos~(SCAC)~\cite{wilkinghoff2021sub}.
These loss functions not only maximize the inter-class distance but also minimize the intra-class compactness, and are known to be effective for the ASD task~\cite{wilkinghoff2023angular}.
Additionally, it has been shown that it achieves better performance with fixed class centers, which prevents trivial projection~\cite{Wilkinghoff2023design,ruff2018deep}.

During training, a data augmentation technique mixup is widely used, and its effectiveness has been well-demonstrated~\cite{JieIESEFPT2023,Wilkinghoff2023design,wilkinghoff2024self,kuroyanagi2022improvement}.
To further improve performance, a new training technique called FeatEx~\cite{wilkinghoff2024self} has been recently proposed.
It trains feature extractors with the following loss:
\begin{align}
    \mathcal{L}_{\rm cat}(\bm{z}^{\rm cat}_i, \bm{l}_i)+\mathcal{L}_{\rm ex}(\bm{z}^{\rm ex}_i, \bm{l}^{\rm ex}_i),\label{eq:featex}
\end{align}
where $\mathcal{L}_{\rm cat}(\cdot, \cdot)$ is a loss function with fixed class centers, $\mathcal{L}_{\rm ex}(\cdot, \cdot)$ is a loss function with trainable centers, $\bm{l}$ is one-hot vectors of labels, and $C$ is the number of classes.
The label is created by simply concatenating the machine type and attribute, and mixup is also used in FeatEx.
$\mathcal{L}_{\rm cat}(\bm{z}^{\rm cat}_i, \bm{l}_i)$ is the original loss and $\mathcal{L}_{\rm ex}(\bm{z}^{\rm ex}_i, \bm{l}^{\rm ex}_i)$ is the additional loss introduced by FeatEx.
$\bm{z}^{\rm ex}_i$ and $\bm{l}^{\rm ex}_i$ are obtained using randomly selected $i$-th and $j$-th samples as follows:
\begin{align}
\vspace{-3pt}
    \bm{z}^{\rm ex}_i&=[\bm{z}^{(1)}_i, \bm{z}^{(2)}_j]\in\mathbb{R}^{2D},\\
    \bm{l}^{\rm ex}_i &= 
    \begin{cases}
        [\bm{l}_i, \bm{0}, \bm{0}]\in[0,1]^{3C}, & \text{if } i = j \\
        [\bm{0}, 0.5\cdot\bm{l}_i, 0.5\cdot\bm{l}_j]\in[0,1]^{3C}, & \text{if } i \neq j 
    \end{cases}
\vspace{-3pt}
\end{align}
where $\bm{0}$ is a $C$-dimensional zero vector.
The additional loss term $\mathcal{L}_{\rm ex}(\bm{z}^{\rm ex}_i, \bm{l}^{\rm ex}_i)$ has been interpreted as encouraging the feature extractor to identify whether $\bm{z}^{(1)}_i$ and $\bm{z}^{(2)}_j$ belong to the same audio signal, thereby capturing more information in a self-supervised learning~(SSL) manner~\cite{wilkinghoff2024self}.

\subsection{Backend}
\label{sec:backend}
The backend is responsible for calculating the anomaly scores.
It is constructed using the features $\bm{z}^{\rm cat}$ of the training data.
As a preliminary step, it obtains the cluster centers by applying the k-means clustering to the features of the source domain.
Then, an anomaly score is calculated by the smallest distance from an observed feature to the cluster centers of the source domain and all features of the target domain.

\section{Proposed method}
\label{sec:proposed}
In this section, according to the DCASE Challenge Task2~\cite{Nishida_arXiv2024_01}, we consider a scenario where only machine-type labels are available, while attribute labels are not.
This is a reasonable scenario because different machines are typically monitored with different sensors, and machine-type labels can be easily obtained from the sensor index.
In contrast, annotating attribute labels requires additional monitoring costs.

\subsection{Enhanced feature extractor}
The performance of the discriminative approach is degraded without the attribute labels~\cite{wilkinghoff2023angular}.
We mitigate this problem by enhancing the feature extractor.

\subsubsection{Multi-resolution spectrograms}
\label{sec:pr_ml}
We extend the conventional feature extractor by adding an amplitude spectrogram of the different resolution to the input features (i.e., the number of input features increases from $2$ to $M$).
We expect that the multi-resolution spectrograms capture anomalies from different perspectives, leading to better performance.

\subsubsection{Subspace loss}
\label{sec:pr_sl}
We propose a subspace loss to achieve nearly the same performance improvement effect as FeatEx but in a simpler way.
First, $\mathcal{L}_{\rm ex}(\cdot, \cdot)$ in~\eqref{eq:featex} can also be interpreted as it encourages each network $g_m(\cdot)$ to identify the labels from the corresponding input features $f_m(\bm{x}_i)$ without other features $\bm{z}^{\rm cat}_i$.
We then use the following loss function, replacing $\mathcal{L}_{\rm ex}(\cdot, \cdot)$ with subspace loss functions $\mathcal{L}_m(\cdot, \cdot)$.
\begin{align}
    \mathcal{L}_{\rm cat}(\bm{z}^{\rm cat}_i, \bm{l}_i)+\sum_{m=1}^M\mathcal{L}_m(\bm{z}^{(m)}_i, \bm{l}_i),
\end{align}
where $\mathcal{L}_m(\cdot, \cdot)$ is a loss function with trainable centers.
The subspace loss is more parameter-efficient than FeatEx with respect to $M$. 
For the additional loss terms, FeatEx requires $DCM(M+1)$ parameters whereas subspace loss requires only $DCM$ parameters.
In this respect, the subspace loss is well-suited for using multi-resolution spectrograms.

\subsection{Pseudo-labeling}
\label{sec:pr_pattr}
Next, we improve the performance using the pseudo-labeling. 
Our proposed pseudo-labeling methods consist of two steps: 1) construction of a feature space and 2) obtaining pseudo-labels by clustering within the feature space.

\subsubsection{Construction of the feature space} 
We propose three approaches for constructing the feature space.

\noindent
\textbf{Classification of available labels (Class):}
This method trains the feature extractors $g_m(\cdot)$ with only available labels, and utilizes the extracted features $\bm{z}^{\rm cat}$ for the clustering step.

\noindent
\textbf{External pre-trained model:}
This method constructs the feature space using external pre-trained models.
We employ PANNs~\cite{kong2020panns} and OpenL3~\cite{cramer2019look,arandjelovic2017look} because they have been used for ASD~\cite{Wilkinghoff2023pretrained,dohi2024distributed}.
PANNs are models pre-trained on supervised audio event classification tasks, while OpenL3 is a model pre-trained on a SSL task that predicts correspondence between an audio clip and a video clip.
Both models are trained on the large-scale dataset, Audioset~\cite{gemmeke2017audio}.
We use CNN-14 from PANNs and an OpenL3 model trained on an environmental subset.

\noindent
\textbf{Triplet learning (Triplet):}
In preliminary experiments, we found that the external pre-trained models can reflect differences in noise.
This motivates us to construct a noise-robust feature space from scratch.
In this method, a feature extractor is trained with the triplet loss~\cite{schroff2015facenet} using $\bm{X}^{\rm anc}$, $\bm{X}^{\rm pos}$, $\bm{X}^{\rm neg}$ as anchor, positive, and negative samples, respectively.
\begin{align}
\label{eq:triplet1}
\bm{X}^{\rm anc}_i &\sim \{\bm{X}_i, \mathrm{Noise}(\bm{X}_i)\} \quad \text{with equal probability},\\
\label{eq:triplet2}
\bm{X}^{\rm pos}_i &= \mathrm{Noise}(\bm{X}_i),\\
\label{eq:triplet3}
\bm{X}^{\rm neg}_i &\sim \{\mathrm{Resize}(\bm{X}_i), \mathrm{Noise}(\mathrm{Resize}(\bm{X}_i)), \bm{X}_j, \notag \\ &\quad \mathrm{Noise}(\bm{X}_j)\} \quad \text{with equal probability},
\end{align}
where $\bm{X}_i$ and $\bm{X}_j$ are amplitude spectrograms of different signals from the same machine type, $\mathrm{Noise}(\cdot)$ adds sound from different machine types, and $\mathrm{Resize}(\cdot)$ resizes the spectrogram.
$\mathrm{Noise}(\cdot)$ helps ignore noise and the use of $\bm{X}^{\rm neg}$ helps capture differences in machine sounds.

\subsubsection{Obtaining pseudo-attribute labels by clustering}
\label{sec:pr_cl}
We apply Gaussian mixture models~(GMMs)-based clustering to obtain pseudo-labels.
Because the dimension of the feature is relatively high compared to the number of samples, we reduce the number of dimension to two by UMAP~\cite{mcinnes2018umap} and then apply GMM-based clustering.
The number of clusters is determined by the Bayesian information criterion~(BIC).

\section{Experimental evaluations}
\subsection{Experimental setups}
We conducted experimental evaluations using the DCASE 2023~\cite{Dohi_arXiv2023_01} and 2024~\cite{Nishida_arXiv2024_01} Task~2 Challenge dataset (ToyADMOS2~\cite{Harada2021}, MIMII DG~\cite{Dohi2022}).
Both datasets consist of development (\texttt{dev}) and evaluation (\texttt{eval}) subsets with different machine types.
The 2023 dataset included seven machines each for \texttt{dev} and \texttt{eval} sets, and the 2024 dataset included seven machines for the \texttt{dev} set and nine machines for the \texttt{eval} set.
The 2023 dataset included attribute labels, while the 2024 dataset did not include them for three machines in the \texttt{dev} set and four machines in the \texttt{eval} set.
The training data included 1,000 samples of normal data for each machine type, of which 990 samples are in the source domain and 10 samples are in the target domain.
The test data included 100 samples of normal or anomalous sounds for each domain of each machine type.
Each recording was a $6$ to $18$-second single channel signal sampled at \SI{16}{\kHz}.

In STFT, the frame shift was half of the DFT size, and frequency bins in the range of \SIrange{200}{8000}{\Hz} were used.
In Triplet, the DFT size of STFT was $1024$, the signal-to-noise ratio of $\mathrm{Noise}(\cdot)$ was randomly selected from $[-5, 5)$, the scale of $\mathrm{Resize}(\cdot)$ was randomly selected from $[0.5, 0.8)$ or $[1.2, 1.5)$, and the margin parameter of the triplet loss was $1$.
In the GMM-based clustering described in Sec.~\ref{sec:pr_cl}, we set the maximum number of clusters to $16$ for the source domain and $2$ for the target domain, respectively.

We trained the feature extractors $g_m(\cdot)$ for $16$ epochs with $64$ of batch size and the feature extractor of Triplet for $6$ epochs with $32$ of batch size, respectively.
The optimizer was AdamW~\cite{adamw2019} with a fixed learning rate of $0.001$.
The feature extractors consisted of the convolutional networks similar to that in~\cite{wilkinghoff2024self}.
The mixup and FeatEx were applied with \SI{50}{\percent} probability.
We used SCAC as the loss function where its number of sub-clusters was set to $16$ and its scale parameter was fixed.
During inference, we used the conventional backend described in Sec.~\ref{sec:backend} with $16$ clusters for the source domain.
We averaged anomaly scores for each audio signal using scores obtained from checkpoints of $12$, $14$, and $16$ epochs.

As the metrics, we used DCASE Challenge Task~2 official metrics: harmonic mean of the area under the curve~(AUC) and partial AUC with $p=0.1$ over all machine types and domains.
We calculated the arithmetic mean and standard deviation of the official scores across five independent trials.

\begin{table}[t]
    \caption{Evalution of the multi-resolution spectrograms. 
    The values are presented as ``mean (standard deviation)''.}
    \label{tbl:exp_ml}
    \vspace{-5pt}
    \centering
    \scalebox{0.88}{
    \begin{tabular}{l|llll}
        \toprule
        DFT size of STFT & \texttt{23dev} & \texttt{23eval} & \texttt{24dev} & \texttt{24eval} \\
        \midrule
        256 & 63.97 (0.77) & 62.03 (1.96) & 61.62 (1.42) & 51.95 (1.46) \\
        1024 & 63.92 (0.39) & 63.32 (1.14) & 60.44 (0.42) & 53.71 (0.33) \\
        4096 & \textbf{66.83} (1.01) & 61.53 (1.35) & 58.92 (0.72) & 52.70 (1.10) \\
        256, 1024 & 64.67 (0.88) & \textbf{65.75} (0.62) & \textbf{62.36} (0.43) & 54.37 (0.65) \\
        256, 4096 & 66.78 (0.77) & 64.38 (0.92) & 61.95 (0.78) & 54.24 (0.47) \\
        1024, 4096 & 66.79 (0.49) & 63.60 (1.41) & 60.64 (0.86) & \textbf{55.71} (0.49) \\
        256, 1024, 4096 & 66.81 (0.82) & 64.78 (0.71) & 61.35 (0.74) & 54.33 (0.96) \\
        \bottomrule
        \end{tabular}
    }
    \vspace{-11pt}
\end{table}

\begin{table}[t]
    \caption{Evalution of the subspace loss.
    The values are presented as ``mean (standard deviation)''.}
    \label{tbl:exp_sl}
    \vspace{-5pt}
    \centering
    \scalebox{0.92}{
        \begin{tabular}{l|llll}
            \toprule
             & \texttt{23dev} & \texttt{23eval} & \texttt{24dev} & \texttt{24eval} \\
            \midrule
            N/A & 66.78 (0.77) & 64.38 (0.92) & 61.95 (0.78) & 54.24 (0.47) \\
            FeatEx & \textbf{67.52} (1.36) & \textbf{69.42} (0.93) & 63.30 (0.81) & 55.20 (0.63) \\
            Subspace loss & 67.21 (0.66) & 68.83 (0.94) & \textbf{64.08} (0.40) & \textbf{56.95} (0.92) \\
        \bottomrule
        \end{tabular}
    }
    \vspace{-12pt}
\end{table}

\subsection{Effectiveness of the enhanced feature extractor}
First, we evaluated the performance of using single or multiple-resolution spectrograms.
We always used a spectrum as one of the input features and did not use FeatEx and subspace loss.
We conducted experiments under both unlabeled (DCASE 2024) and labeled conditions (DCASE 2023).
Table~\ref{tbl:exp_ml} summarizes the evaluation results, where we can see that the combined use of multiple-resolution spectrograms is basically effective in all datasets although desirable resolution differs.
The best performance is basically achieved by combining the two high-performing cases, indicating that the original performance is inherited even when multiple-resolution spectrograms are combined.

Next, we evaluated the performance of using the subspace loss.
We used a spectrum and the multi-resolution spectrograms with DFT sizes of $256$ and $4096$ as input features, i.e., $M=3$.
Table~\ref{tbl:exp_sl} summarizes the evaluation results.
It can be seen that not only FeatEx but also the subspace loss further improves the performance across all datasets.
Although the better method depends on the dataset, the subspace loss improves the performance in a simpler and more parameter-efficient manner.

\begin{table*}[t]
    \caption{Evalution of the pseudo-labeling methods.
    The values are presented as ``mean (standard deviation)''. \\
    N/A does not use pseudo-labels and GT uses ground truth labels.}
    \label{tbl:exp_pattr}
    \vspace{-5pt}
    \centering
    \scalebox{0.88}{
    \begin{tabular}{llllllll|l}
        \toprule
        \texttt{23dev} & bearing & fan & gearbox & slider & ToyCar & ToyTrain & valve & hmean \\
        \midrule
        N/A & 59.77 (2.19) & 52.21 (3.83) & \textbf{72.87} (1.44) & \textbf{96.90} (0.79) & 44.35 (2.68) & 51.81 (1.61) & 74.18 (3.28) & 60.65 (1.43) \\
        Class & 65.32 (1.16) & 58.30 (2.92) & 67.48 (2.19) & 92.32 (1.93) & 47.32 (1.01) & 49.94 (1.72) & 61.91 (6.72) & 60.53 (0.61) \\
        Triplet & 65.82 (1.23) & 58.00 (1.69) & 67.64 (2.05) & 89.60 (2.00) & 50.04 (2.04) & 51.88 (1.30) & \textbf{82.83} (6.36) & \textbf{63.74} (0.80) \\
        PANNs & 65.45 (1.37) & \textbf{63.53} (2.39) & 70.00 (0.45) & 85.62 (2.14) & 49.99 (0.72) & 51.84 (1.61) & 70.01 (1.22) & 63.29 (0.56) \\
        OpenL3 & \textbf{67.22} (0.53) & 62.14 (0.90) & 66.15 (1.38) & 88.81 (1.97) & \textbf{53.68} (1.05) & \textbf{53.98} (0.82) & 63.98 (1.94) & 63.54 (0.37) \\
        \midrule
        GT & 66.90 (1.00) & 60.99 (1.51) & 71.75 (1.57) & 85.41 (2.44) & \textbf{56.44} (1.23) & \textbf{59.24} (1.46) & 79.93 (3.67) & \textbf{67.21} (0.66) \\
        \midrule
        \midrule
        \texttt{23eval} & Vacuum & ToyTank & ToyNscale & ToyDrone & bandsaw & grinder & shaker & hmean \\
        \midrule
        N/A & 63.72 (3.05) & 54.57 (1.75) & 65.54 (1.28) & 49.83 (2.15) & 44.83 (2.92) & 59.94 (2.07) & 44.02 (12.26) & 52.82 (3.27) \\
        Class & 77.27 (4.46) & 58.55 (3.40) & 66.05 (7.55) & 48.64 (1.27) & 54.10 (1.59) & 58.44 (2.47) & 53.72 (7.67) & 58.10 (1.21) \\
        Triplet & 76.91 (3.95) & 64.94 (2.45) & 74.87 (3.92) & 53.00 (1.63) & 54.89 (1.85) & 57.53 (1.28) & 53.52 (2.46) & 60.84 (0.31) \\
        PANNs & \textbf{85.29} (3.17) & 65.11 (1.52) & \textbf{77.44} (1.77) & \textbf{54.14} (1.66) & 53.39 (1.16) & 58.28 (0.60) & 51.05 (3.65) & 61.37 (0.74) \\
        OpenL3 & 78.30 (1.71) & \textbf{65.23} (1.94) & 73.14 (1.58) & 50.74 (1.22) & \textbf{61.54} (1.85) & \textbf{62.55} (0.81) & \textbf{76.32} (2.36) & \textbf{65.50} (0.51) \\
        \midrule
        GT & \textbf{89.14} (0.63) & 63.20 (1.92) & \textbf{86.03} (1.75) & \textbf{57.34} (1.08) & 58.08 (0.72) & 60.93 (2.17) & \textbf{83.97} (4.68) & \textbf{68.83} (0.94) \\
        \midrule
        \midrule
        \texttt{24w/o} & gearbox & slider & ToyTrain & AirCompressor & BrushlessMotor & HoveringDrone & ToothBrush & hmean \\
        \midrule
        N/A & 63.00 (1.36) & 82.57 (3.23) & 61.10 (1.05) & 55.52 (1.38) & 51.83 (1.07) & 58.30 (3.40) & \textbf{56.24} (2.83) & 59.98 (0.41) \\
        Class & 62.85 (1.12) & 84.91 (7.19) & \textbf{61.97} (1.02) & 62.87 (1.32) & 54.99 (1.36) & 58.99 (2.36) & 54.67 (4.47) & \textbf{61.79} (1.88) \\
        Triplet & 61.26 (1.22) & \textbf{90.03} (0.96) & 61.42 (1.16) & 53.73 (4.67) & \textbf{55.21} (2.11) & 60.36 (2.59) & 46.08 (4.64) & 58.86 (2.18) \\
        PANNs & \textbf{70.17} (0.69) & 83.78 (2.28) & 61.06 (2.24) & 54.98 (1.39) & 54.23 (1.61) & 60.00 (2.03) & 43.96 (3.54) & 58.93 (0.74) \\
        OpenL3 & 62.53 (2.17) & 75.97 (2.62) & 60.31 (1.31) & \textbf{63.96} (1.83) & 53.64 (1.59) & \textbf{62.09} (1.06) & 46.30 (2.04) & 59.43 (0.99) \\
        \midrule
        GT & \textbf{78.78} (2.52) & 88.83 (1.40) & \textbf{65.45} (0.52) & \textbf{69.66} (0.84) & 51.29 (3.37) & \textbf{66.74} (1.78) & 49.52 (2.30) & \textbf{64.59} (0.99) \\
        \bottomrule
        \end{tabular}
    }
        \vspace{-12pt}
\end{table*}

\begin{figure}[t]
    \centering
    \includegraphics[width=0.99\columnwidth]{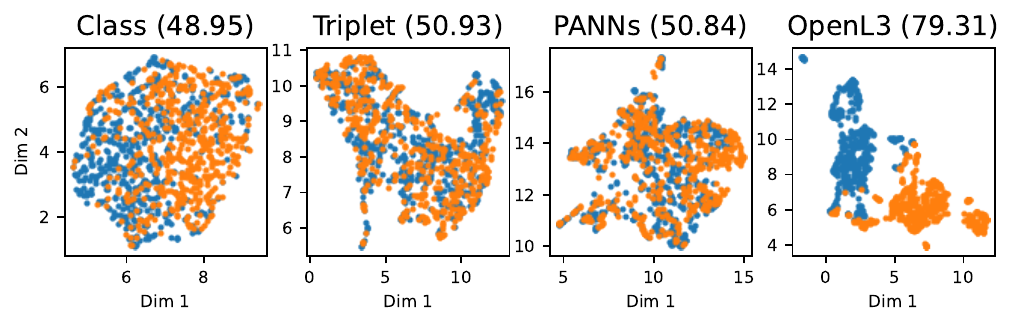}
    \vspace{-20pt}
    \caption{
    Visualizations of the feature space for the shaker in the source domain of the 2023 dataset, colored by ground-truth labels.
    The value in the parentheses represents the official score.
    This is from one of five trials.
    }
    \vspace{-10pt}
    \label{fig:shaker}
  \end{figure}

\begin{figure}[t]
    \centering
    \includegraphics[width=0.99\columnwidth]{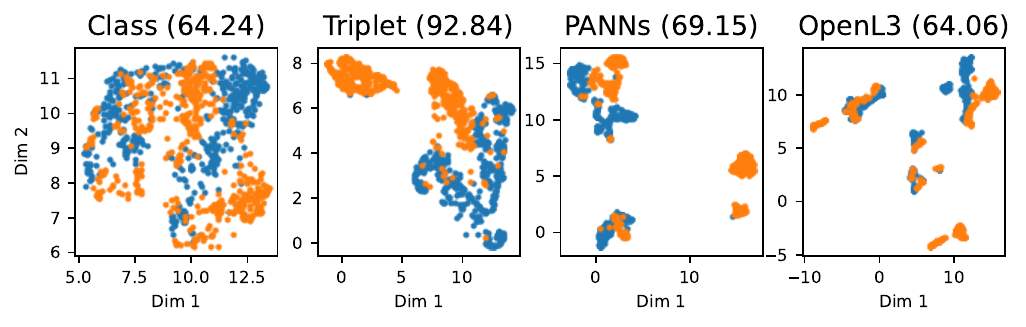}
    \vspace{-20pt}
    \caption{Visualizations of the feature space for the valve in the source domain of the 2023 dataset, colored by ground-truth labels.
    The value in the parentheses represents the official score.
    This is from one of five trials.
    }
    \label{fig:valve}
    \vspace{-15pt}
  \end{figure}

\subsection{Effectiveness of the pseudo-labeling}
\label{sec:exp_pl}
Finally, we compared the performance using several pseudo-labeling methods.
We trained the feature extractors $g_m(\cdot)$ with the subspace loss, using the spectrum and multi-resolution spectrograms with DFT sizes of $256$ and $4096$ as input features.
In addition to the 2024 dataset, we conducted experiments using the 2023 dataset with the concealed attribute labels for all machines.

Table~\ref{tbl:exp_pattr} summarizes the evaluation results, where \texttt{24w/o} indicate the machines without attribute labels in the 2024 dataset.
For machines with the attribute labels in the 2024 dataset, the harmonic mean of the scores for N/A, Class, Triplet, PANNs, OpenL3, and GT were $59.78$, $59.37$, $59.36$, $59.56$, $58.57$, and $59.52$, respectively, indicating that differences in the pseudo-labels could slightly affect the performance of the machine with labels.
From the table, we can see that the ground truth labels are extremely effective in most cases, and all of the proposed pseudo-labeling methods also significantly improve the performance, including an absolute improvement of \SI{30}{\percent} (shaker in \texttt{23eval}).
Although the best pseudo-labeling method varies by machine type, the external pre-trained models basically achieve better performance.
In the ToothBrush of \texttt{24w/o}, even the ground-truth labels are not effective, and we observed that Class achieves similar performance to N/A by constructing only one cluster (i.e., a meaningless but harmless feature space).

Figure~\ref{fig:shaker} shows that OpenL3 successfully captures the differences in machine sounds, and the generated pseudo-labels significantly improve the performance.
In contrast, Fig.~\ref{fig:valve} shows that the external pre-trained models do not capture the differences in machine sounds, and the generated pseudo-labels degrade performance compared to N/A. 
Upon analyzing the feature spaces, we observed that the external pre-trained models tended to form clusters based on the noise differences, which explains why the pseudo-labels degrade performance.
Even in such cases, Triplet generates effective pseudo-labels by constructing a noise-robust feature space from scratch.

\section{Conclusion}
In this paper, we investigated ways to improve the performance of discriminative methods under the unlabeled conditions.
First, to achieve better performance in the unlabeled conditions, we enhanced the feature extractor using multi-resolution spectrograms and the subspace loss.
Second, we proposed several pseudo-labeling methods to effectively train the feature extractor, including Class, Triplet, and the use of external pre-trained models.
Our experimental results showed that 
1) both multi-resolution spectrograms and the subspace loss improved the performance with and without labels,
2) all of the proposed pseudo-labeling methods could improve the performance in most cases,
3) the external pre-trained models especially achieved better performance,
and 4) Triplet is more effective in noisy conditions.



\section*{References}
\printbibliography




\end{document}